\documentstyle[preprint,aps]{revtex}
\input epsf.tex
\def\DESepsf(#1 width #2){\epsfxsize=#2 \epsfbox{#1}}
\begin{document}
\preprint{\vbox{\hbox{UM-P-95/117, OITS-596}}}
\draft
\title{$B$ Decays And Models For CP Violation}
\author{Xiao-Gang He}
\address{
Research Center for High Energy Physics\\
School of Physics,
University of Melbourne\\
Parkville, Vic 3052, Australia\\
and\\
Institute of Theoretical Science\\
University of Oregon\\
Eugene, OR 97403-5203, USA}
\date{December 1995}
\maketitle
\begin{abstract}
The decay modes $B$ to $\pi\pi$, $\psi K_S$, $K^- D$, $\pi K$ and 
$\eta K$ are promising channels to study the unitarity
triangle of the CP violating CKM matrix. In this paper I study the
consequences of these measurements in the Weinberg model. 
I show that using the same set of measurements, the following 
different mechanisms for CP violation can be distinguished:
1) CP is violated in the CKM sector
only; 2) CP is violated spontaneously in the Higgs sector only; And
3) CP is violated in both the CKM and Higgs sectors.
\end{abstract}
\pacs{}
\section{Introduction}
CP violation is one of the unresolved mysteries in 
particle physics. The explanation in the Standard Model (SM) based on 
Cabibbo-Kobayashi-Maskawa (CKM) matrix\cite{1} is still not established,
although there is no conflict between the observation of CP violation 
in the neutral K-system\cite{2} and theory\cite{3}, intriguing hints of other
plausible explanations emerge from  consideration of baryon 
asymmetry of the universe\cite{4}. Models based on additional Higgs bosons\cite{5,6}
can equally well explain the existing laboratory 
data\cite{chen} and provide large CP violation
required from baryon asymmetry\cite{4}.
It is important to carry out more 
experiments to find out the origin of CP violation. It is for this reason 
that exploration of CP violation in $B$ system is so crucial. 
The $B$ system offers 
several final states that provide a rich source for the study of this 
phenomena\cite{7}. Several methods using $B$ decay modes have been proposed to
measure the phase angles, $\alpha = Arg(-V_{td}V_{tb}^*/V_{ub}^*V_{ud})$, 
$\beta=Arg(-V_{cd}V_{cb}^*
/V_{tb}^*V_{td})$ and $\gamma=Arg(-V_{ud}V_{ub}^*/V_{cb}^*V_{cd})$ in 
the unitarity triangle of the CKM matrix\cite{8,9,10,11,12,13}. 
It has been shown that $\bar B^0 (B^-)\rightarrow \pi^+\pi^-, \pi^0\pi^0 
(\pi^-\pi^0)$\cite{10}, $\bar B^0\rightarrow \psi K_S$\cite{11} 
and $B^-\rightarrow K^- D$\cite{12} decays can be used to 
determine
$\alpha $, $\beta$ and $\gamma$, 
respectively.  Recently 
it has been shown that 
$B^-\rightarrow \pi^-\bar K^0,\;\pi^0 K^-,\; \eta K^-$ and 
$B^-\rightarrow \pi^-\pi^0$ can also be used to determine $\gamma$\cite{13}.  
If the sum of these
three angles is $180^0$, the SM is a good model for CP
violation. Otherwise new mechanism for CP violation is needed.
In this paper  I study the
consequences of these measurements  in the 
Weinberg model.

In the Weinberg model CP can be violated in the CKM 
sector and Higgs sector. If CP is violated spontaneously, it 
occurs in the
Higgs sector only. 
I will call the model with CP violation in both the CKM and Higgs
sectors as WM-I, and the model with CP violation only in the Higgs sector
as WM-II. 
There are many ways to distinguish the SM and 
Weinberg model for CP 
violation. For example the neutron electric dipole moment in the Weinberg model 
is several orders of magnitude larger than the SM prediction\cite{14}. However, 
the neutron electric dipole moment measurement alone 
can not distinguish the WM-I from the  WM-II.
I show that measurements of CP violation in $B$ decays not only can  be used to 
distinguish the SM from  the Weinberg model, but can also be used
to determine  whether CP is violated in the Higgs sector only 
 or in both the CKM and Higgs 
sectors.  \\

\noindent
{\bf $B$ Decay Amplitudes In The SM}

CP violation in the SM is due to the phase in the CKM mixing matrix in
the charged current interaction,
\begin{eqnarray}
L = -{g\over 2\sqrt{2}}\bar U\gamma^\mu(1-\gamma_5)V_{KM} D W^+_\mu + H.C.\;,
\end{eqnarray}
where $U = (u,\;c,\;t)$, and $D = (d,\;s,\;b)$. $V_{KM}$ is the CKM
matrix. For three generations, it is a $3\times 3$ unitary matrix. It has three
rotation angles and one non-removable phase 
which is the source of CP violation in the SM. 
I will use the Maiani, 
Wolfenstein and Chau-Keung\cite{wck} convention for the CKM
matrix, in which $V_{ub}^*$ has the phase $\gamma$, and $V_{td}^*$ has the 
phase $\beta$ and other CKM elements have no or very small phases.

The effective Hamiltonian responsible 
for $\Delta C = 0$ hadronic $B\rightarrow \pi\pi,\; \pi K,\; \eta K\;, 
\psi K_S$ 
decays at the quark level to 
the one loop level in 
electroweak interaction can be parameterized as,
\begin{eqnarray}
H_{eff} &=& {G_F\over \sqrt{2}}[V_{ub}V_{uq}^*(c_1O_{1u}^q+c_2O_{2u}^q)
+V_{cb}V_{cq}^*(c_1O_{1c}^q+c_2O_{2c}^q)\nonumber\\
&-&
\sum_{i=3}^{12}[V_{ub}V_{uq}^* c^u_i +  V_{cb}V_{cq}^*c^c_i +  
V_{tb}V_{tq}^*c^t_i]O_i^q\;,
\end{eqnarray}
where $c_i^f$ (f = u, c, t) 
are Wilson Coefficients (WC) of the corresponding quark 
and gluon operators $O_i^q$.
The superscript $f$ indicates the internal quarks. 
$q$ can be $d$ or $s$ quark depending on if the decay 
is a $\Delta S = 0$ or $\Delta S = -1$ process.
The operators $O_i^q$ are defined as
\begin{eqnarray}
 &O&^{q}_{1f} = \bar q_{\alpha} \gamma_{\mu} L 
f_{\beta} \bar f_{\beta} \gamma^{\mu} L b_{\alpha} ,
  \ \  O^{q}_{2f} = \bar q \gamma_{\mu} L f 
\bar f \gamma^{\mu} L b ,           \nonumber \\ 
 &O&_{3(5)} = \bar q \gamma_{\mu} L b \Sigma 
\bar q^{\prime} \gamma^{\mu} L(R) q^{\prime} ,   
  \ \ O_{4(6)} = \bar q_{\alpha} \gamma_{\mu} L 
b_{\beta} \Sigma \bar q^{\prime}_{\beta} 
       \gamma^{\mu} L(R) q^{\prime}_{\alpha}  ,  
                       \nonumber \\ 
 &O&_{7(9)} = {3 \over 2} \bar q \gamma_{\mu} L 
b \Sigma e_{q^{\prime}} \bar q^{\prime} 
       \gamma^{\mu} R(L) q^{\prime} ,        
  \ \ O_{8(10)} ={3 \over 2} \bar q_{\alpha} 
\gamma_{\mu} L b_{\beta} \Sigma e_{q^{\prime}} 
       \bar q^{\prime}_{\beta} \gamma^{\mu} R(L) 
q^{\prime}_{\alpha} ,      \nonumber \\ 
&O&_{11} ={g_{s}\over{32\pi^2}}m_{b}\bar q \sigma_{\mu \nu} RT_{a}b 
G_{a}^{\mu \nu} \;,\;\;
Q_{12} = {e\over{32\pi^2}} m_{b}\bar q \sigma_{\mu \nu}R b 
F^{\mu \nu} \;,
\end{eqnarray} 
where $L(R) = (1 \mp \gamma_5)$, 
and $q^{\prime}$ is summed over $u$, $d$, 
$s$, and $c$ quarks.  The subscripts $\alpha$ and $\beta$ are 
the color indices.  
$T^{a}$ is the color SU(3) generator with the normalization 
$Tr(T^{a} T^{b}) = \delta^{ab}/2$.  
$G^{\mu \nu}_{a}$ and $F_{\mu \nu}$ are the gluon 
and photon field strengths, respectively. 
$O_1$, $O_2$ are the tree level and QCD 
corrected operators.  $O_{3-6}$ are the gluon 
induced strong penguin operators.  
$O_{7-10}$ are the electroweak penguin operators due to $\gamma$ and 
$Z$ exchange, and ``box'' diagrams at 
loop level. The WC's $c_{1-10}$ have been evaluated at the next-to-leading-log 
QCD corrections 
\cite{15}. The operators $O_{11,12}$ are the dipole 
penguin operators. Their
WC's have been evaluated at the leading order in QCD correction\cite{16},
and their phenomenological implications in $B$ 
decays have also been studied\cite{17}.
 
One can generically parameterize the decay amplitude of $B$ as
\begin{eqnarray}
\bar A_{SM} = <final\;state|H_{eff}^q|B> = V_{fb}V^*_{fq} T(q)^f 
+ V_{tb}V^*_{tq}P(q)\;,
\end{eqnarray}
where $T(q)$ contains the $tree$ and $penguin$ due to internal 
u and c quark contributions, while $P(q)$ contains $penguin$
contributions from internal t and c or u quarks. 
I use $\bar A$ for the decay amplitude of $B$ meson containing a 
b quark, and $A$ for a $B$ meson containing a $\bar b$ quark. 
The WC's involved in $T$ are
much larger than the ones in $P$. 
One expects the hadronic matrix elements arising from quark 
operators to be the same order of magnitudes. The relative 
strength of the amplitudes $T$ and $P$ are
predominantly determined by their corresponding WC's in 
the effective Hamiltonian. In general $|P|$, if not zero, 
is about or less than 10\% of $|T|$.

For $\bar B^0\rightarrow \psi K_S$, the decay amplitude can be written as
\begin{eqnarray}
\bar A_{SM}(\psi K) &=& V_{cb}V_{cs}^* T_{\psi K} + V_{tb}V_{ts}^* P_{\psi K}\nonumber\\
&=& |V_{cb}V_{cs}|(T_{\psi K}-P_{\psi K}) +
|V_{ub}V_{us}^*|e^{-i\gamma}P_{\psi K}\;.
\end{eqnarray} 
The second term is about $10^3$ times
smaller than the first term and can be safely neglected. To this level,
the decay amplitude for $\bar B^0\rightarrow \psi K_S$ does not contain
 weak CP violating phase. This decay mode provides a clean way to measure
the phase angle $\beta$ in the SM\cite{11}.

One can parameterize the decay amplitudes for $B\rightarrow \pi\pi\;,\;
K\pi\;,\; \eta K$ in a similar way. Further if flavor SU(3) symmetry is a good 
symmetry there are certain relations among the decay 
amplitudes\cite{18}. I will assume the validity of the
SU(3) symmetry in my later analysis. 
The operators $Q_{1,2}^u$, $O_{1,2}^c$, 
$O_{3-6, 11,12}$, and $O_{7-10}$ transform under SU(3)
symmetry as $\bar 3_a + \bar 3_b +6 + \overline {15}$,
$\bar 3$, 
$\bar 3$, and $\bar 3_a + \bar 3_b +6 + \overline {15}$, respectively.
Flavor SU(3) symmetry predicts
\begin{eqnarray}
\sqrt{2}\bar A( \pi^0\pi^0) + 
\sqrt{2}\bar A( \pi^-\pi^0) = \bar 
A(\pi^+\pi^-)\;,\label{pp}\\
\sqrt{2} \bar A(\pi^0 K^-) 
- 2\bar A(\pi^- \bar K^0) 
= \sqrt{6} \bar A(\eta_8 K^-)\;.
\label{pk}
\end{eqnarray}
Isospin symmetry also imply eq.(\ref{pp}).
These relations form two triangles in the complex plan which 
provide important information for obtaining phase angles $\alpha$ and
$\gamma$\cite{10,13}.

I parameterize the decay amplitudes in the SM as
\begin{eqnarray}
\bar A_{SM}(\pi^-\pi^0) &=& |V_{ub}V_{ud}^*|e^{-i\gamma}T_{\pi^-\pi^0}
+ |V_{tb}V_{td}^*|e^{i\beta}P_{\pi^-\pi^0}\;,\nonumber\\
\bar A_{SM}(\pi^+\pi^-) &=&|V_{ub}V_{ud}^*|e^{-i\gamma}T_{\pi^+\pi^-}
+ |V_{tb}V_{td}^*|e^{i\beta}P_{\pi^+\pi^-}\;,\nonumber\\
\bar A_{SM}(K^-\pi^0) &=& |V_{ub}V_{us}^*|e^{-i\gamma}T_{K^-\pi^0}
+ |V_{tb}V_{ts}^*|P_{K^-\pi^0}\;,\nonumber\\
\bar A_{SM}(\bar K^0\pi^-)&=& |V_{ub}V_{us}^*|e^{-i\gamma}T_{\bar K^0\pi^-}
+ |V_{tb}V_{ts}^*|P_{\bar K^0\pi^-}\;,\label{sm}
\end{eqnarray}
The decay amplitudes $\bar A_{SM}(\pi^0\pi^0)$ and $\bar A_{SM}(K^-\eta)$ are
obtained by the SU(3) relations in eqs.(\ref{pp}) and (\ref{pk}).
 
I would like to point out that 
$\bar A_{SM}(\pi^-\pi^0)$ and $\sqrt{2}\bar A(K^-\pi^0)-\bar A(\bar K^0 \pi^-)$ 
only receive contributions from the effective operators which transform 
as $\overline{15}$\cite{13,19},
\begin{eqnarray}
&&\bar A_{SM}(\pi^-\pi^0) = V_{ub}V_{ud}^*C_{\overline{15}}^T + V_{tb}V_{td}^*
C_{\overline{15}}^P\;,\nonumber\\
&&\bar A_{SM}(K^-\pi^0)-\bar A_{SM}(\bar K^0\pi^-)/\sqrt{2} 
= V_{ub}V_{us}^*C_{\overline{15}}^T + V_{tb}V_{ts}^*
C_{\overline{15}}^P\;,\label{re}
\end{eqnarray}
where $C_{\overline{15}}$ is the invariant amplitude due to operators
transform as $\overline {15}$ under SU(3) symmetry.
This is an important property useful for my later discussions. 
The second term in $\bar A_{SM}(\pi^-\pi^0)$ is less than
3\% of the first term\cite{dh1}. For all practical purposes it can be neglected.
However, the second term on the right hand side of the second equation
in eq.(\ref{re}) can not be neglected because there is an enhancement factor 
$|V_{tb}V_{ts}^*|/|V_{ub}V_{us}^*| $ which is about 50\cite{19}.

The effective Hamiltonian responsible for $B\rightarrow DK$ decay
is given by
\begin{eqnarray}
H_{eff} &=& {G_F\over \sqrt{2}}[V_{ub}V_{cs}^*(c_1\bar u^\alpha\gamma_\mu L 
b_\beta 
\bar s^\beta \gamma^\mu L c_\alpha + c_2\bar u\gamma_\mu L b 
\bar s \gamma^\mu L c)\nonumber\\
&+&V_{cb}V_{us}^*(c_1\bar c^\alpha\gamma_\mu L 
b_\beta 
\bar s^\beta \gamma^\mu L u_\alpha +c_2\bar c\gamma_\mu L b 
\bar s \gamma^\mu L u)]\;.
\end{eqnarray}
The decay amplitudes for $B^-\rightarrow K^-D^0$ and
$B^-\rightarrow K^-\bar D^0$, respectively, are given by 
\begin{eqnarray}
\bar A_{SM}(K^-D^0) &=& |V_{ub}V_{cs}^*|a_{KD}e^{-i\gamma}\;,\nonumber\\
\bar A_{SM}(K^-\bar D^0) &=& |V_{cb}V_{us}^*|b_{KD}\;.
\end{eqnarray}
From the above, one easily obtains the decay amplitude for 
$B^-\rightarrow K^- D_{CP}$ with $D_{CP} = (D^0-\bar D^0)/\sqrt{2}$ being
the CP even eigenstate,
\begin{eqnarray}
\bar A_{SM}(K^-D_{CP}) = 
{1\over \sqrt{2}}[\bar A_{SM}(K^-D^0) - \bar A_{SM}(K^-\bar D^0)]\;.
\label{kd}
\end{eqnarray}
This relation form a triangle in the complex plan which is useful
in determining the phase angle $\gamma$ in the SM\cite{12}.\\ 

\noindent
{\bf $B$ Decay Amplitudes In The Weinberg Model}

 In the
Weinberg model, besides the CP violating phase in the CKM matrix, CP 
violation for hadronic
$B$ decays can also arise from the exchange of charged Higgs at tree and loop
 levels, and also neutral Higgs at  loop levels. In this model, there are two 
physical charged Higgs particles and three neutral Higgs particles. 
The neutral Higgs couplings to fermions are flavor conserving and proportional
to the fermion masses. Flavor changing decay amplitude can only be generated 
at loop level. For the cases in consideration, all involve light
fermions, the CP violating amplitude generated by neutral Higgs exchange 
is
very small and can be neglected. 
The exchange of charged Higgs may generate sizable CP violating
decay amplitudes, however. 
The charged Higgs couplings to fermions are given by\cite{21}
\begin{eqnarray}
L = 2^{7/4}G_F^{1/2}\bar U [V_{KM}M_D(\alpha_1H_1^+ + \alpha_2 H^+)R
+M_UV_{KM}(\beta_1 H^+_1 +\beta_2 H_2^+)L]D + H.C.\;,
\end{eqnarray}
where $M_{U,D}$ are the diagonal up and down quark mass matrices.
The parameters $\alpha_i$ and $\beta_i$ are obtained from diagonalizing charged 
Higgs masses and can be written as, 
\begin{eqnarray}
\alpha_1 &=& s_1c_3/c_1\;,\;\;\alpha_2 = s_1s_3/c_1\;,\nonumber\\
\beta_1 &=& (c_1c_2c_3+s_2s_3e^{i\delta_H})/s_1c_2\;,\;\;
\beta_2 = (c_1c_2s_3-s_2c_3e^{i\delta_H})/s_1c_2\;,
\end{eqnarray}
where $s_i = sin\theta_i$ and $c_i = cos\theta_i$ with $\theta_i$ being the 
 rotation
angles, and $\delta_H$ is a CP violating phase. The decay amplitudes 
due to exchange of charged Higgs at tree level will be proportional to 
$V_{fb}V_{f'q}^*(m_b m_{f'}/m_{H_i}^2)\alpha_i \beta_i^*$. Therefore if a
decay involves light quark,  the amplitude will be suppressed. However, at one 
loop level if the internal quark masses are large, sizable CP violating
decay amplitude may be generated. The leading term is from the strong dipole
penguin interaction with top quark in the loop\cite{22},
\begin{eqnarray}
L_{DP} &=& \tilde f O_{11}\;,\nonumber\\
\tilde f &=& {G_F\over 16\sqrt{2}}\sum_i^2 \alpha^*_i\beta_i
V_{tb}V_{tq}^* {m_t^2\over m_{H_i}^2-m_t^2}(
{m_{H_i}^4\over (m_{H_i}^2-m_t^2)^2} \mbox{ln}{m_{H_i}^2\over m_t^2}
-{m_{H_i}^2\over m_{H_i}^2-m_t^2} - {1\over 2})\;.
\end{eqnarray}
This is not suppressed compared with the penguin contributions in the SM. 
There is also a similar contribution 
from the operator $O_{12}$. However the WC of this operator is 
suppressed by a factor of $\alpha_{em}/\alpha_s$ 
and its contribution to $B$ decays can be neglected.
I write the $O_{11}$ contribution to $B$ decays as
\begin{eqnarray}
\bar A_{final } = V_{tb}V_{tq}^*a_{final}e^{i\alpha_H}\;,
\end{eqnarray}
where $\alpha_H$ is the phase in $\tilde f$ which is decay mode independent,
and $a_{final} = |\tilde f|<final \;state|O_{11}|B>$ which is
decay mode dependent. Note that $L_{DP}$ transforms as $\bar 3$ under SU(3)
symmetry. It does not contribute to 
$\bar A(\pi^-\pi^0)$ and $\sqrt{2} \bar A( K^-\pi^0) - \bar A(\bar K^0\pi^-)$. 

The decay amplitudes in the Weinberg model can be written as 
\begin{eqnarray}
\bar A_W(\pi^+\pi^-) &=& \bar A_{SM}(\pi^+\pi^-)+ 
V_{tb}V_{td}^*e^{i\alpha_H}a_{\pi\pi}\;,
\nonumber\\
\bar A_W(\pi^-\pi^0) &=& \bar A_{SM}(\pi^-\pi^0)\;,\nonumber\\
\bar A_W(K^-\pi^0) &=&\bar A_{SM}(K^-\pi^0) + 
{1\over \sqrt{2}}V_{tb}V_{ts}^*e^{i\alpha_H}
a_{K\pi}\;,\nonumber\\
\bar A_W(\bar K^0\pi^-) &=& \bar A_{SM}(\bar K^0\pi^-) + V_{tb}V_{ts}^* 
e^{i\alpha_H}a_{K\pi}\;,\nonumber\\
\bar A_W(\psi K_S) &=& \bar A_{SM}(\psi K_S) + 
V_{tb}V_{ts}^* e^{i\alpha_H} a_{\psi K}\;.\label{wm}
\end{eqnarray}
In the SU(3) limit $a_{\pi\pi} = a_{K\pi}$. 

The decay amplitudes for 
$B\rightarrow
KD$ only have contributions from tree operators. Because the CP 
violating amplitude from tree level charged Higgs exchange is negligibly small, to a
good approximation, 
\begin{eqnarray}
\bar A_W(KD) = \bar A_{SM}(KD)\;.
\end{eqnarray}

The decay amplitudes for both the WM-I and WM-II have the same form given in 
eqs. (17) and (18). In the WM-I CP is violated in both the CKM and Higgs 
sectors with
$\alpha\beta\gamma\alpha_H\ne 0$. In the WM-II CP is violated only in the Higgs
sector with $\alpha=\beta=\gamma=0$, 
but $\alpha_H\ne 0$. I will drop the asterisk of
the CKM matrix elements in the WM-II.

\section{CP violation in B decays}

\noindent
{\bf  $B\rightarrow \pi\pi$ Decays}

In the time evolution of the rate asymmetry for $\bar B^0\rightarrow \pi^+\pi^-$
and $B^0\rightarrow \pi^-\pi^+$, there are two terms varying with time, one
varies as a cosine function and the other as a sine function. The coefficients 
of these two terms can be measured experimentally. The coefficient
of the sine term is given by\cite{11}
\begin{eqnarray} 
 Im\lambda =Im\left ({q \over p} {\bar A(\pi^+\pi^-) \over A(\pi^-\pi^+)}\right )\;,
\end{eqnarray} 
where $p$ and $q$ are the mixing parameters defined by
\begin{eqnarray}
|B_H> = p|\bar B^0> +q|B^0>\;,\;\;|B_L> = q|\bar B^0> - p|B^0>\;,
\end{eqnarray}
where $|B_{H,L}>$ are the heavy and light mass eigenstates, respectively.

In the SM, the mixing is dominated by the top quark loop in the box diagram,
and 
\begin{eqnarray}
{q\over p} = {V_{tb}^*V_{td}\over V_{tb}V_{td}^*} = e^{-2i\beta}\;.
\end{eqnarray}
One obtains
\begin{eqnarray}
 Im \lambda &=& Im\left ( e^{-2i(\beta+\gamma)} {e^{i\gamma}\bar A_{SM}(\pi^+\pi^-)
\over e^{-i\gamma} A_{SM}(\pi^-\pi^+)} \right )\nonumber\\
&=&Im \left (e^{2i\alpha} {|V_{ub}V_{ud}^*|T_{\pi^+\pi^-}
+|V_{tb}V_{td}^*|P_{\pi^+\pi^-}e^{i(\beta+\gamma)}\over 
|V_{ub}V_{ud}^*|T_{\pi^+\pi^-}
+|V_{tb}V_{td}^*|P_{\pi^+\pi^-}e^{-i(\beta+\gamma)}}\right ) \nonumber\\
&=& {|\bar A_{SM}(\pi^+\pi^-)|
\over | A_{SM}(\pi^-\pi^+)|}sin (2\alpha+\theta_{+-})\;. \label{tri}
\end{eqnarray}
The ratio $|\bar A_{SM}|/|A_{SM}|$ can be determined from time integrated
rate asymmetry at symmetric\cite{dh} and asymmetric colliders\cite{7}.
If $\theta_{+-}$ can be determined, 
the phase angle $\alpha$ can be determined.
To determine $\theta_{+-}$, Gronau and London\cite{10} 
proposed to use the isospin relation in eq.(\ref{pp}), 
\begin{eqnarray}
\sqrt{2}\bar A( \pi^0\pi^0) + 
\sqrt{2}\bar A( \pi^-\pi^0) = \bar 
A(\pi^+\pi^-)\;,
\end{eqnarray}
and normalize
the amplitudes $\bar A_2 = \sqrt{2} e^{i\gamma}A_{SM}(\pi^-\pi^0)$ and 
$A_2 = \sqrt{2} e^{-i\gamma}A_{SM}(\pi^+\pi^0)$ on the real axis.
The triangle is shown in Figure 1.
It is easy to see from eq.(22) 
that the angle $\theta_{+-}$ is given by phase angle difference between 
$\bar A_1 = e^{i\gamma}\bar A_{SM}(\pi^+\pi^-)$ and $A_1 =
e^{-i\gamma}A_{SM}(\pi^-\pi^+)$. It can be easily read off from Figure 1. 

In the Weinberg model, similar measurement will  obtain different
result. In  the 
WM-I, in addition to the phase $\beta$, there is also a phase $\beta_H$ in $q/p$ due to 
charged Higgs exchange in the box diagram. One obtains 
$q/p = e^{-2i(\beta+\beta_H)}$, and
\begin{eqnarray}
Im\lambda &=& Im\left (e^{-2i(\beta +\gamma +\beta_H)} { \bar A_{SM}(\pi^+\pi^-) + V_{tb}V_{td}^*
e^{i(\alpha_{H}+\gamma)}a_{\pi\pi}\over
A_{SM}(\pi^+\pi^-) + V_{tb}V_{td}^*
e^{-i(\alpha_{H}+\gamma)}a_{\pi\pi}}\right )\nonumber\\
&=& {|\bar A_{W}(\pi^+\pi^-)|\over |A_W(\pi^-\pi^+)|}sin (2\alpha-2\beta_H+\theta_{+-}^H)\;.
\end{eqnarray}
This equation has the same form as eq.(\ref{tri}) 
for the SM. The determination of
 $\alpha-\beta_H$ is exactly the same as $\alpha$ in the 
SM except  that in this case 
$\bar A_1 = e^{i\gamma}\bar A_W(\pi^+\pi^-)$ and 
$A_1 = e^{-i\gamma}A_W(\pi^-\pi^+)$. 
The phase $\beta_H$ can be neglected because it is suppressed by a factor of
$m_bm_t/m_H^2$. 
The measurement proposed here still measures $\alpha$ even though there is an 
additional 
CP violating phase in $\bar B^0\rightarrow \pi^+\pi^- (\pi^0\pi^0)$ decay 
amplitudes. 

If CP is violated spontaneously, the result will be dramatically different. 
Here the CP violating weak phases in $\bar A_{SM}$ are all zero.
The amplitude 
$\bar A_2 = \sqrt{2}\bar A_W(\pi^-\pi^0)$ is equal to $A_2 = \sqrt{2}A_W(\pi^+\pi^0)$, and can be normalized to be real. 
Now using the isospin triangle
in Figure 1, one easily obtains the phase in $\bar A_W(\pi^+\pi^-)/A_W(\pi^-\pi^+)$,
and therefore determine the phase angle $\beta_H$. 
One would obtain a very small value. This will
be a test for spontaneous CP violation model WM-II.

Using the isospin triangle in Figure 1, 
the CP violating amplitude $|a_{\pi\pi}|^2sin^2\alpha_H$ in the WM-II can also be
determined. It is given by
\begin{eqnarray}
|a_H|^2sin^2\alpha_H = {L^2\over 4|V_{tb}V_{td}|^2} \label{cpa}
\end{eqnarray}
This measurement will also serve as a test for  the WM-II. I 
will come back to this later.\\

\noindent
{\bf  $B\rightarrow \psi K_S$ Decay}

In the SM, the cleanest way to measure
$\beta$ is to measure the parameter $Im\lambda$ 
for $\bar B^0\rightarrow
\psi K_S$ decay \cite{11}. In this case,
\begin{eqnarray}
Im\lambda_{\psi K} = Im \left ({q\over p}
{\bar A_{SM}(\psi K_S)\over  A_{SM}(\psi K_S)}\right )\;.
\end{eqnarray}
Neglecting the small term proportional to $V_{cb}V_{cs}^*$, 
one obtains
\begin{eqnarray}
Im \lambda_{\psi K} = Im\left ({q\over p}{V_{cb}V_{cs}^*
\over V_{cb}^*V_{cs}}\right) 
= -sin(2\beta)\;.
\end{eqnarray} 
This is a very clean way to measure the phase angle $\beta$ in the SM.

In the Weinberg model, the same measurement will
give different result. In the WM-I, one has
\begin{eqnarray}
Im\lambda_{\psi K} = Im\left(e^{-2i(\beta+\beta_H)} {\bar A_{SM}(\psi K_S) + V_{tb}V_{ts}^*
e^{i\alpha_H}a_{\psi K} \over A_{SM}(\psi K_S) + V_{tb}V_{ts}^*
e^{-i\alpha_H}a_{\psi K}}\right )\;.
\end{eqnarray}
The amplitude from the new  contribution proportional to $a_{\psi K}$ 
is expected to be about 10\% of the SM 
contribution.
Even though $\beta_H$ is small, $Im\lambda_{\psi K}$ in the WM-I will be different 
from $-sin(2\beta)$.  This measurement alone will not
 be able to distinguish the SM and WM-I. 
However combining the result from this measurement and knowledge about $\alpha$ 
determined from the previous section and $\gamma$ to be determined in the next section, one can distinguish the SM from the WM-I. 
If the SM is the correct model, 
the phase angles $\alpha$, $\gamma$ and $\beta$ will 
add up to $180^o$. However if the WM-I is the right model and one naively 
interprets
$Im\lambda_{\psi K_S}$ to be $-sin(2\beta)$, the sum of the three phase angles
 will not be $180^o$. 

In the WM-II, the measurement $Im\lambda_{\psi K_S}$ will only measure the phase 
difference between $\bar A(\psi K_S)$ and $A(\psi K_S)$ which is smaller than 
the value for the SM and WM-I. A small experimental
value  for $Im\lambda_{\psi K}$ is an indication for the WM-II.\\

\noindent
{\bf $B^-\rightarrow K^- D$ Decays}

In the SM and WM-I, the triangle relation 
\begin{eqnarray}
A(K^- D^0)-A( K^-\bar D^0) = \sqrt{2}A(K^-D_{CP})\;,
\end{eqnarray}
provides a measurement for the phase angle $\gamma$.  
The phase $\gamma$ is given as shown
in Figure 2\cite{12}. 

In the WM-II, there are no CP violating weak phase
angles in these decay amplitudes. 
The triangles for the particle decays and anti-particle decays
will be identical. 
One should be aware that in the SM and WM-I if the strong
rescattering phases are all zero the triangles
for the particle and anti-particle decays will also be identical, one
must put the two triangles on the opposite side as shown in Figure 2
to determine the value for $\gamma$. 
However, if the
two triangles for the particle and anti-particle decays are not identical,
the WM-II is ruled out.\\ 

\noindent
{\bf $B^-\rightarrow \pi K\;,\;\;\eta K$ Decays}

Another method to measure the phase angle $\gamma$ is to use 
the following $B$ decays: $B^{-} \rightarrow \pi^{-} \bar K^{0}$, 
$\pi^{0} K^{-}$, $\eta K^{-}$, and 
$B^{-} \rightarrow \pi^{-} \pi^{0}$\cite{13} .  This method requires 
the construction of the triangle mentioned in eq.(\ref{pk})
\begin{eqnarray} 
 \sqrt{2} \bar A(K^- \pi^{0} ) 
-2 \bar A( \bar K^{0}\pi^-) 
    &=& \sqrt{6} \bar A(K^{-}\eta_8). 
\end{eqnarray}
   In the spectator model, 
the contributions to $\bar A(\bar K^0\pi^-)$ from the tree operators
vanish\cite{wolf}. To a good approximation, one has
\begin{eqnarray} 
 \bar A_{SM}( \bar K^{0}\pi^-) 
= A_{SM}( K^{0}\pi^+)\;. 
\end{eqnarray}
These amplitudes do not have weak phases. 
They can be  normalized to be real. From Figure 3, one can determine
the two amplitudes $B$ and $\bar B$ which are given by
\begin{eqnarray} 
 B &=& \sqrt{2} \bar A_{SM}(K^-\pi^0 ) 
- \bar A_{SM}( \bar K^0 \pi^-) , 
      \nonumber \\ 
 \bar B &=& \sqrt{2} A_{SM}(K^{+}\pi^0) 
      - A_{SM}( K^0 \pi^{+} ).
\end{eqnarray}
Using SU(3) relation in eq.(\ref{re}), one obtains
\begin{eqnarray} 
 B - \bar B = -i 2 \sqrt{2} e^{i \delta^{T}} {|V_{us}| \over |V_{ud}|} 
   |\bar A_{SM}(\pi^{-} \pi^{0})| 
\sin{\gamma} ,                         \label{BB} 
\end{eqnarray}  
The angle $\delta^{T}$ denotes the 
strong final state rescattering phase of 
the tree amplitude of $B$ (or $\bar B$).  It is clear that 
$\mbox{sin}\gamma$ can be determined\cite{13}. 

In the WM-I, the result will be different. In this case even in the spectator model, 
$\bar A_W(\bar K^0\pi^-)$ is not equal to $A_W(K^0\pi^+)$ because the new contribution proportional to
$V_{tb}V_{ts}^*a_{K\pi}e^{i\alpha_H}$. There is no common
side for the triangles for the particle and anti-particle decay amplitudes.  
No  useful information about the phase $\gamma$ can be obtained.  
However if experiments will find 
$\bar A(\bar K^0\pi^-) \ne A(K^0\pi^+)$, it indicates 
that the SM may not be correct.

In the WM-II, the analysis is again very different. In the analysis for the SM, 
 the decay amplitudes for $\bar A( \bar K^0 \pi^-)$ and $A_{SM}(K^0\pi^+)$
are normalize to be real.   In the WM-II 
because the  additional term
$V_{tb}V_{ts}e^{i\alpha_H}a_{K\pi}$, one can no longer 
use this normalization.
One needs to find amplitudes which can serve as the orientation axis. 
To this end  I note that the amplitude $B$ and $\bar B$ in the above
only receive contributions from operators in
the effective Hamiltonian transforming
as $\overline {15}$ under SU(3). The strong dipole penguin, which
transforms as $\bar 3$, does not contribute, and therefore in the WM-II,
$B = \bar B$. One can normalize the triangles by putting
$B$ and $\bar B$ on the real axis as shown in Figure 4.
The phases of the rest decay amplitudes can be easily read off from the 
figure. One particularly interesting amplitude is
$|a_{K\pi}|^2sin^2\alpha_H$.
From Figure 4, one obtains
\begin{eqnarray}
|a_{K\pi}|^2 sin^2\alpha_H = {L'^2\over 4 |V_{tb}V_{ts}^*|^2}\;.
\end{eqnarray}

Several comments on these measurements are in order:
1) In the SM, $\bar A(\bar K^0\pi^-) = A(K^0\pi^+)$. This is not generally true 
in the Weinberg model. This can be used to test the SM.
2) In the WM-II, $B =\bar B$. This is a test for WM-II. In the SM and WM-I
this happens only when the strong rescattering phases are zero which 
is, however,  unlikely.
3) If SU(3) is a good symmetry, the quantity   $|a_{K\pi}|^2sin^2\alpha_H$ 
is equal to $|a_{\pi\pi}|^2
sin^2\alpha_H$ obtained in eq.(\ref{cpa}). 
This will serve as a test for the WM-II.
\\

\noindent
{\bf Rate differences in $\bar B^0\rightarrow \pi^+\pi^-$ and
$\bar B^0\rightarrow K^- \pi^+$}

In this section I comment on rate differences in $\bar B^0\rightarrow
\pi^+\pi^-$ and $\bar B^0\rightarrow
K^-\pi^+$. 
 
The decay amplitude for $\bar B^0\rightarrow K^-\pi^+$ can be written
as
\begin{eqnarray}
A(K^-\pi^+) = V_{ub}V_{us}^*T_{K^-\pi^+} +  
V_{tb}V_{ts}^*(P_{K^-\pi^+} + {1\over \sqrt{2}} e^{i\alpha_H}a_{K\pi})\;.
\end{eqnarray}
It has been shown that in the SU(3) limit, $T(P)_{\pi^+\pi^-}=
T(P)_{K^-\pi^+}$, and $a_{\pi\pi} = a_{K\pi}$\cite{19}. Here $T(P)_{\pi^+\pi^-}$ 
and $a_{\pi\pi}$ are the corresponding amplitudes in 
$\bar A_{SM}(\pi^+\pi^-)$ given in eq.(\ref{sm}). In the SM, 
\begin{eqnarray}
\Delta_{\pi\pi} &=& \Gamma(\pi^+\pi^-) - \bar \Gamma(\pi^-\pi^+)
= -Im(V_{ub}V_{ud}^*V_{tb}V_{td}^*)
Im(T_{\pi^+\pi^-}P^*_{\pi^+\pi^-}) {m_B\Lambda_{\pi\pi}\over 4\pi}\;,
\nonumber\\
\Delta_{K\pi} &=& \Gamma(K^- \pi^+) - \bar \Gamma(K^+ \pi^-)
= -Im(V_{ub}V_{ud}^*V_{tb}V_{td}^*)
Im(T_{\pi^+ K^-}P^*_{\pi^+ K^-}) {m_B\Lambda_{\pi K}\over 4\pi}\;,
\end{eqnarray}
where $\Lambda_{ab} = \sqrt{1-2(m_a^2+m_b^2)/m_B^2 + (m_a^2-m_b^2)^2/m_B^4}$. One obtains\cite{19}
\begin{eqnarray}
{\Delta(\pi^+\pi^-) \over \Delta (\pi^+ K^-)} = -1\;.
\end{eqnarray}
When SU(3) breaking effects are included, 
\begin{eqnarray}
{\Delta(\pi^+\pi^-)\over \Delta(\pi^+ K^-)} \approx -{f_\pi^2\over f_K^2}\;.
\end{eqnarray}
The ratio is negative and of order one. In the Weinberg model, 
the prediction is very different. 
In the WM-I, the situation is complicated. It is difficult to obtain useful information about  
CP violating parameters. In the simpler case, the WM-II, 
\begin{eqnarray}
{\Delta(\pi^+\pi^-)\over \Delta(\pi^+K^-)}
 = {V_{td}\over V_{ts}}{V_{ub} V_{ud}Im(T_{\pi^+\pi^-}a_H^*)+
V_{tb}V_{td} Im (P_{\pi^+\pi^-} a_H^*)\over
V_{ub}V_{us}Im(T_{\pi^+K^-}a_H^*)+
V_{tb}V_{ts} Im (P_{\pi^+K^-} a_H^*)}\;.
\end{eqnarray}
This is very different from the SM prediction. The ratio can vary a large range. 
If the first term dominates, 
$\Delta (\pi^+\pi^-)/\Delta(\pi^+ K^-) = V_{td}V_{ud}/V_{ts}V_{us}$ which 
is of order one. If the second term dominates, the ratio is given by $V_{td}^2/V_{ts}^2$ which is positive, but very small. 

\section {Conclusion}

I have analyzed the consequences of several methods for measuring CP violating
observables. 
These measurements can be used to distinguish different models for CP violation. 

The isospin triangle relation among $\bar B^0\rightarrow \pi^+\pi^-,\; \pi^0\pi^0$ and
$B^-\rightarrow \pi^-\pi^0$ provides critical information to correct strong penguin 
contamination in the determination of the phase angle $\alpha$. The same measurement also
determine the phase angle $\alpha$ in the WM-I where CP is violated both in the CKM and 
Higgs sectors. 
In the WM-II where CP is violated spontaneously in the Higgs sector, 
one would obtain a very
different  value. The WM-II will be tested. 

The asymmetry in time evolution for $\bar B^0\rightarrow \psi K_S$ is an idea place to determine 
$\beta$ in the SM. This is, however, not true for the WM-I because the 
contamination from the Higgs induced strong dipole penguin operator. This measurement does not 
measure the true value of $\beta$ in this case.  
In the WM-II 
the resulting $Im \lambda_{\psi K}$ is very small.
This measurement again provides a test for the WM-II.

The triangle relation among $B^-\rightarrow K^- \bar D^0,\; K^0 D^0,\;K^- D_{CP}^0$ provides a clean 
way to measure the phase angle $\gamma$ in both the SM and WM-I. Combining these measurement and 
the previous measurements for $\alpha$ and $\beta$, 
it is possible to distinguish the SM and WM-I because in the SM the true 
$\beta$ is measured in $\bar B^0\rightarrow \psi K_S$, whereas in the WM-I 
the measurement is contaminated. 
If the sum of the three phase angles is $180^o$, the SM is the correct one. If the WM-II 
is the correct model, experiments will find the triangles for particle and 
anti-particle decays to be 
identical. 

The method to measure $\gamma$ using $B\rightarrow K \pi,\; K\eta,\; \pi\pi$ will provide different 
results for the three models. In principle the three models of CP violation can be distinguished. 
This analysis is based on a triangle relation obtained in SU(3) symmetry. The validity of SU(3) 
flavor symmetry has not been established. One should be careful when carry out this analysis. 
SU(3) breaking effects may change the
results. Detailed study is needed.

Another way to distinguish models for CP violation is to use the rate asymmetry
$\Delta(\pi^+\pi^-)$ and $\Delta(\pi^+ K^-)$. The SM and Weinberg model have
very different predictions.

 I would like to thank Professors Gunion and Han for hospitality at the University of California, Davis where part of this work was done. 
This work was supported in part by the Department of Energy Grant 
No. DE-FG06-85ER40224 and  by Australian Research Council.

\begin{figure}[htb]
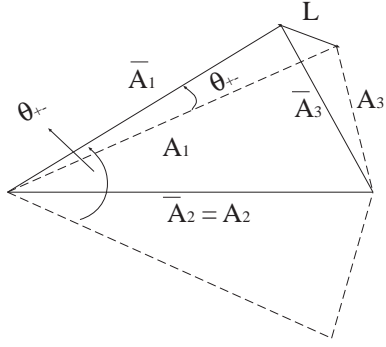

\centerline{ \DESepsf(triangle1.epsf width 6 cm) }
\smallskip
\caption {The isospin triangle for $B\rightarrow \pi\pi$. In the SM and WM-I, 
$\bar A_1 = e^{i\gamma} \bar A_{SM, W}(\pi^+\pi^-)$, $\bar A_2 =\protect\sqrt{2} 
e^{i\gamma}\bar A_{SM, W}(\pi^-\pi^0)$,
$\bar A_3 =\protect\sqrt{2} e^{i\gamma}\bar A_{SM, W}(\pi^0\pi^0)$,  and 
$A_1 = e^{-i\gamma} A_{SM, W}(\pi^-\pi^+)$, $ A_2 = \protect\sqrt{2} e^{-i\gamma} 
A_{SM,W}(\pi^+\pi^0)$,
$ A_3 = \protect\sqrt{2} e^{-i\gamma}\bar A_{SM,W}(\pi^0\pi^0)$.
 In the WM-II, the same as for
the WM-I except that all weak phases  $\alpha$, $\beta$ and $\gamma$ are zero.}

\end{figure}

\begin{figure}[htb]
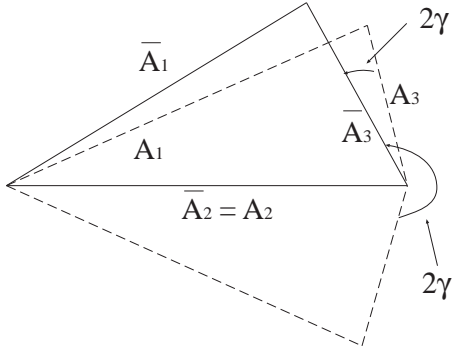

\centerline{ \DESepsf(triangle2.epsf width 6 cm) }
\smallskip
\caption {The triangle relation for $B\rightarrow K D$. 
$\bar A_1 = \protect\sqrt{2} \bar A_{SM,W}(K^-D_{CP})$, 
$\bar A_2 = \bar A_{SM,W}(K^- \bar D^0)$,  $\bar A_3 = \bar A_{SM,W}(K^- D^0)$, 
and $ A_1 = \protect\sqrt{2} 
A_{SM,W}(K^+D_{CP})$, $A_2 =  A_{SM,W}(K^+ D^0)$,  $ A_3 = A_{SM,W}(K^+ \bar D^0)$}.
\end{figure}

\begin{figure}[htb]
\centerline{ \DESepsf(triangle3.epsf width 6 cm) }
\smallskip
\caption {The  triangle relation for $B\rightarrow K\pi, \;K \eta$ in the SM. 
$\bar A_1 =\protect\sqrt{2} \bar  A_{SM}(K^-\pi^0)$, $\bar A_2 = 2\bar A_{SM}(\bar K^0 \pi^-)$,
$\bar A_3 =\protect\sqrt{6} \bar A_{SM}(K^-\eta_8)$, and $A_1 =\protect\sqrt{2} A _{SM}(K^+\pi^0)$, 
$ A_2 = 2 A_{SM}(K^0 \pi^+)$,
$\bar A_3 =\protect \sqrt{6} \bar A_{SM}(K^+\eta_8)$.}
\end{figure}

\begin{figure}[htb]
\centerline{ \DESepsf(triangle4.epsf width 6 cm) }
\smallskip
\caption {The  triangle relation for $B\rightarrow K\pi, \;K \eta$ in the WM-II. 
$\bar A_1 =\protect\sqrt{2} \bar  A_{W}(\bar K^0\pi^-)$, $\bar A_2 = 2\bar A_W(K^-\pi^0)$,
$\bar A_3 =\protect\sqrt{6} \bar A_{W}(K^-\eta_8)$, and $A_1 =\protect\sqrt{2} A _{W}( K^0\pi^+)$, 
$ A_2 = 2A_W(K^+\pi^0)$,
$ A_3 =\protect\sqrt{6}  A_{SM}(K^+\eta_8)$.}
\end{figure}

\end{document}